\documentclass[%
 reprint,
 amsmath,amssymb,
 aps,
]{revtex4-2}

\usepackage{float}
\usepackage{graphicx}
\usepackage{float}
\usepackage[utf8]{inputenc}
\usepackage{textgreek}
\usepackage{dcolumn}
\usepackage{xcolor}
\usepackage{bm}
\usepackage{bbm}
\usepackage{subcaption}
\usepackage{hyperref}
\hypersetup{
    colorlinks=true,       
    linkcolor=blue,        
    citecolor=blue,      
    urlcolor=blue}
    
\begin{document}

\preprint{APS/123-QED}

\title{Neutrino Phenomenology and Leptogenesis from Non-Invertible Fusion Symmetry}

\author{Pulakesh Borah}
 \email{Electronic adress: \textcolor{magenta}{pborah651@gmail.com}}

\author{Mrinal Kumar Das}
 \email{Electronic adress: \textcolor{magenta}{mkdas@tezu.ernet.in}}
\affiliation{
 Department of Physics, Tezpur University, Napaam, Assam, India-784028}

\begin{abstract}

We constructed a minimal type-I seesaw model with two right-handed neutrinos, in which the flavor structure is determined by a non-invertible \(\mathbb{Z}_2\)-gauged \(\mathbb{Z}_5\) fusion symmetry~($\mathbb{Z}^{IN}_5$). The framework, incorporating a second Higgs doublet and an auxiliary \(\mathbb{Z}_2\) parity, yields predictive neutrino mass textures without introducing flavon fields. A fit to neutrino data selects the inverted mass ordering with \(m_3=0\) and predicts a nearly maximal Dirac CP phase, together with a characteristic prediction for the effective Majorana mass governing neutrinoless double-beta decay($0\nu\beta\beta$). We further investigate thermal leptogenesis and show that the framework can reproduce the observed baryon asymmetry of the Universe. Our results establish a connection between predictive neutrino phenomenology and leptogenesis within a minimal framework based on non-invertible fusion symmetry.

\end{abstract}

\maketitle

\section{\label{sec:level1}Introduction}

The origin of flavour remains among the least understood aspects of the Standard Model (SM). Fermion masses and mixing angles span several orders of magnitude, and within the SM, the Yukawa couplings responsible for this hierarchy are free parameters fixed only by experiment. The earliest attempts to explain this hierarchy introduced a horizontal continuous symmetry, most notably the Froggatt-Nielsen mechanism~\cite{froggatt1979hierarchy}, in which Yukawa couplings are generated as powers of a small symmetry-breaking ratio. This approach was later extended to non-Abelian discrete flavor symmetries such as $A_4$, $S_4$, and $A_5$~\cite{ma2006neutrino,altarelli2009revisiting,altarelli2010discrete,altarelli2006tri,ma2001softly,das2020phenomenological,pathak2026matter,borah2026gravitational,kumar2025neutrino,everett2009icosahedral,ding2012golden}, where specific mixing patterns follow from the representation theory of the chosen group, typically realized through an extended flavon sector and a vacuum-alignment mechanism. More recently, modular flavor symmetries~\cite{feruglio2019neutrino,pathak2026matter} have replaced this flavon sector with a single complex modulus $\tau$, whose vacuum expectation value (VEV) alone breaks the flavor symmetry and fixes the Yukawa couplings as modular forms.

While oscillation measurements confirm that neutrinos have mass and undergo mixing, fundamental parameters like the absolute mass scale, ordering, and CP-violating phases remain unresolved \cite{esteban2024nufit,nufit61}. Furthermore, oscillation frameworks cannot distinguish whether neutrinos are Dirac or Majorana fermions. The primary experimental probe for addressing this is neutrinoless double beta decay ($0\nu\beta\beta$), a lepton-number-violating nuclear process whose detection would confirm the Majorana nature of neutrinos and constrain the effective mass $m_{\beta\beta}$ ~\cite{jones2021physics, dolinski2019neutrinoless, bilenky2012neutrinoless, gomez2011sense}. A Majorana mass term for the neutrino, together with a dynamical explanation of its smallness, is naturally supplied by the type-I seesaw mechanism~\cite{brdar2019type, minkowski1977mu, mohapatra1980neutrino}, in which the exchange of heavy right-handed neutrinos (RHNs) suppresses the light-neutrino mass relative to the electroweak scale. The same heavy-neutrino sector simultaneously offers a dynamical origin for the baryon asymmetry of the Universe (BAU) via leptogenesis~\cite{buchmuller2005leptogenesis, giudice2004towards, vives2006flavor}: the out-of-equilibrium, CP-violating decay of the RHNs generates a lepton asymmetry, which electroweak sphalerons partially convert into a baryon asymmetry, naturally satisfying all three Sakharov conditions~\cite{sakharov1998violation, sakharov1991violation}. Nevertheless, this mechanism comes at the cost of new free parameters, namely RHN masses and Yukawa couplings, mirroring the wider flavor puzzle. To constrain these parameters, flavor symmetries are systematically introduced to the seesaw sector to constrain the RHN and neutrino Yukawa textures alongside the charged leptons.

Most recently, a conceptually distinct organizing principle has emerged for generating such textures: a non-invertible fusion symmetry \cite{kobayashi2025radiative}. Unlike an ordinary group, where the product of two symmetry charges is a single, uniquely determined charge, a non-invertible fusion algebra allows the product of two charges to decompose into a sum of several sectors. An interaction is permitted whenever this fusion product contains the identity sector. Such selection rules have recently been shown to generate textures in neutrino models~\cite{jiang2025texture}, without requiring an extended flavon sector or a vacuum-alignment mechanism.

In this work, we construct a minimal type-I seesaw model using the $\mathbb{Z}_5^{IN}$ non-invertible symmetry, extending the SM by only two right-handed neutrinos, a second $SU(2)_L$ Higgs doublet, and a single auxiliary $\mathbb{Z}_2$ parity. The fusion rules alone fix the charged-lepton, the neutrino Dirac and Majorana sector textures, with no flavon fields or vacuum-alignment mechanism required to generate them. While non-invertible symmetries have recently been applied to generate textures in the neutrino sector~\cite{jiang2025texture}, their consequences for the generation of the baryon asymmetry of the Universe have not previously been explored; this is, to our knowledge, the first study of leptogenesis within a non-invertible fusion-symmetry framework. We confront the resulting texture with current NuFIT 6.1 oscillation data~\cite{nufit61}, and compute its predictions for neutrinoless double beta decay and the baryon asymmetry of the Universe via thermal leptogenesis.

This work is organized as follows. In Section II, we briefly discuss the non-invertible $Z_5^{IN}$ fusion symmetry and its fusion rules. The theoretical framework is outlined in Section III, followed by a numerical analysis of the neutrino sector, neutrinoless double beta 
decay, and leptogenesis in Section IV. Finally, we present our conclusions in Section V.

\section{\label{sec:level2} Non-invertible $\mathbb{Z}_5^{IN}$ fusion symmetry } 
Ordinary discrete flavor symmetries assign fields to group elements, and the product of two group elements is a single, uniquely determined element: $g_a g_b = g_c$. An interaction is allowed only when the corresponding product of charges contains the identity. Non-invertible symmetries relax this: the fusion of two symmetry sectors need not close onto a single sector, but instead decomposes as
\begin{equation}
    a \times b = \sum_c N_{ab}^{\ \ c}\, c, \qquad N_{ab}^{\ \ c} \in \mathbb{Z}_{\geq 0}
\end{equation}
and an operator is permitted whenever the fusion product of the relevant field sectors contains the identity sector. This gives selection rules that are not reducible to ordinary group-representation theory, and can generate texture zeros directly from the fusion structure rather than from an externally imposed flavon alignment.

One such fusion symmetry is $Z_5^{IN}$ which is constructed by gauging the $\mathbb{Z}_2$ automorphism $g^k \to g^{-k}$ of $\mathbb{Z}_5 = \{1,g,g^2,g^3,g^4\}$. This identifies $g^k \sim g^{-k}$ and organizes the five elements into three fusion sectors,
\begin{equation}
    [0] = \{1\}, \qquad [1] = \{g, g^4\}, \qquad [2] = \{g^2, g^3\}
\end{equation}
with fusion rules
\begin{equation}
    [k]\times[k'] = [k+k'] + [k-k'] \pmod 5, \qquad [k] \equiv [-k]
\end{equation}
Explicitly,
\begin{equation}
\label{eq:4}
\begin{aligned}[b]
[0]\times[k] &= [k], \\
[1]\times[1] &= [0]+[2], \\
[1]\times[2] &= [1]+[2], \\
[2]\times[2] &= [0]+[1]   
 \end{aligned}
\end{equation}

Unlike an ordinary $\mathbb{Z}_5$, where $g\times g = g^2$ returns a single element, $[1]\times[1] = [0]+[2]$ has two components: the fusion of two $[1]$-charged fields can proceed through either the identity or the $[2]$ channel. This multi-channel structure is not merely a relabeling of group elements is the origin of the non-invertible selection rules used throughout this construction.

\section{\label{sec:citeref}MODEL FRAMEWORK}
In this work, we constructed a model based on the non-invertible $Z_5^{IN}$ fusion symmetry. To study the type-I seesaw mechanism for neutrino mass generation, we employ the fusion-sector assignments listed in Table I. Under the non-invertible $\mathbb{Z}_5^{IN}$ fusion algebra, the three generations of the left-handed lepton doublet $L$ are assigned individually to the fusion sectors $[0]$, $[1]$, and $[2]$, rather than to a single multiplet, since the sectors $[0],[1],[2]$ are one-dimensional classes rather than representations of a non-Abelian group. The right-handed charged leptons $e_R$, $\mu_R$, and $\tau_R$ transform under the same sectors as their corresponding doublets, $[0]$, $[1]$, and $[2]$ respectively, so that the charged-lepton Yukawa matrix is diagonal by the fusion selection rule alone. We extend the SM by introducing two heavy right-handed Majorana neutrinos $N_1, N_2$, assigned to the sectors $[1]$ and $[2]$. The scalar sector comprises two $SU(2)_L$ Higgs doublets $H_\ell$ and $H_\nu$, assigned to $[0]$ and $[1]$ respectively, together with an auxiliary $\mathbb{Z}_2^\nu$ parity under which $H_\nu$ and $N_{1,2}$ are odd and all other fields are even; this parity prevents $H_\nu$ from generating unwanted charged-lepton couplings while leaving the neutrino Yukawa sector unaffected. Unlike constructions based on non-Abelian discrete groups, no flavon fields are required: the neutrino Yukawa and right-handed neutrino mass textures are fixed directly by the fusion rules of Eq.~\eqref{eq:4}, rather than by a vacuum alignment mechanism. 

\begin{table}[h!]
\centering
\renewcommand{\arraystretch}{1.4}
\small
\setlength{\tabcolsep}{5pt}
\caption{Field content of the model with their $Z_5^{IN}$ charges.}
\begin{tabular}{lccccccc}
\hline\hline
Symmetry & \multicolumn{6}{c}{Field content and charge} \\
\hline
& $L_{e,\mu,\tau}$ & $l_R$ & $N_{1}$ & $N_{2}$ & $H_l$ & $H_\nu$ \\
\hline 
$Z_5^{IN}$ & [0][1][2] & [0][1][2] & [1] & [2] & [0] & [1] \\
\hline
$\mathbb{Z}_2^\nu$ & $+$ & $+$ & $-$ & $-$ &$+$ & $-$\\
\hline\hline
\end{tabular}
\end{table}

Considering the charge assignments of Table I, we construct the Lagrangian term by term, retaining only those operators whose total fusion class contains the identity sector $[0]$ and whose total $\mathbb{Z}_2^\nu$ parity is even.

Considering the assignment rules, the charged lepton sector Lagrangian becomes
\begin{equation}
\mathcal{L}c = y_e \overline{L_e} H_\ell e_R + y_\mu \overline{L_\mu} H_\ell \mu_R + y_\tau \overline{L_\tau} H_\ell \tau_R + \text{h.c.}
\end{equation}
giving the charged lepton mass matrix
\begin{equation}
M_\ell = \frac{v_\ell}{\sqrt{2}}\begin{pmatrix} y_e & 0 & 0 \\ 0 & y_\mu & 0 \\ 0 & 0 & y_\tau \end{pmatrix}
\end{equation}
where $\langle H_\ell^0\rangle = v_\ell/\sqrt{2}$.

The structure of the invariant Dirac Lagrangian involving the heavy right-handed fields $N_{i}$ and lepton fields $L_\alpha$ is given by
\begin{equation}
\begin{aligned}
    \mathcal{L}_D &= a \overline{L_e}\tilde{H}_\nu N_1 + b\overline{L_\mu}\tilde{H}_\nu N_2 +  c\overline{L_\tau}\tilde{H}_\nu N_1 + \\ &d\overline{L_\tau}\tilde{H}_\nu N_2 + \text{h.c.}
\end{aligned}
\end{equation}
giving the Dirac mass matrix of the form
\begin{equation}
M_D = \frac{v_\nu}{\sqrt{2}} Y_\nu = \frac{v_\nu}{\sqrt{2}}\begin{pmatrix} a & 0 \\ 0 & b \\ c & d \end{pmatrix}
\end{equation}
with $\langle H_\nu^0\rangle = v_\nu/\sqrt{2}$.

The heavy Majorana sector Lagrangian takes the following form:
\begin{equation}
\mathcal{L}_N = \frac{1}{2}M_1\overline{N_1^c}N_1 + \frac{1}{2}M_2\overline{N_2^c}N_2 + \text{h.c.}
\end{equation}
and the corresponding heavy Majorana mass matrix is of the form 
\begin{equation}
    M_R = \begin{pmatrix} M_1 & 0 \\ 0 & M_2 \end{pmatrix}
\end{equation}
Here, the fusion symmetry alone fixes $M_R$ to be exactly diagonal.

Integrating out $N_1, N_2$ via the type-I seesaw relation
\begin{equation}
    M_\nu = -M_D M_R^{-1} M_D^T
\end{equation}
gives the light neutrino mass matrix of the form

\begin{equation}
\label{eq:Mnu}
M_\nu = -\frac{v_\nu^2}{2}
\begin{pmatrix}
\dfrac{a^2}{M_1} & 0 & \dfrac{ac}{M_1} \\
0 & \dfrac{b^2}{M_2} & \dfrac{bd}{M_2} \\
\dfrac{ac}{M_1} & \dfrac{bd}{M_2} & \dfrac{c^2}{M_1}+\dfrac{d^2}{M_2}
\end{pmatrix}
\end{equation}
Since $\mathrm{rank}(M_R)=2$, $\mathrm{rank}(M_\nu)\leq 2$, so one light neutrino mass vanishes identically; consistency with current oscillation data selects the inverted-ordering branch, $m_3=0$ ( discussed in Sec. IV).

Since charged lepton mass matrix is diagonal so diagonalising matrix $U_\ell = \mathbb{I}$ exactly, so the PMNS matrix is
\begin{equation}
U_{PMNS} = U_\ell^\dagger U_\nu = U_\nu,
\end{equation}
with $U_\nu$ obtained from the Takagi decomposition $U_\nu^T M_\nu U_\nu = \mathrm{diag}(m_1,m_2,0)$.
\subsection*{Scalar potential}
The renormalizable scalar potential consistent with the fusion symmetry and the auxiliary $\mathbb{Z}_2^\nu$ parity is
\begin{align}
V = {} & m_\ell^2 H_\ell^\dagger H_\ell + m_\nu^2 H_\nu^\dagger H_\nu + \frac{\lambda_\ell}{2}(H_\ell^\dagger H_\ell)^2 + \frac{\lambda_\nu}{2}(H_\nu^\dagger H_\nu)^2 \nonumber \\
& + \lambda_3 (H_\ell^\dagger H_\ell)(H_\nu^\dagger H_\nu) + \lambda_4 |H_\ell^\dagger H_\nu|^2 \nonumber \\
& + \left[\frac{\lambda_5}{2}(H_\ell^\dagger H_\nu)^2 + \text{h.c.}\right].
\end{align}
For real VEVs with $v_\ell^2+v_\nu^2=v^2\approx (246~\mathrm{GeV})^2$, the stationarity conditions is given by 
\begin{equation}
m_\ell^2 + \frac{\lambda_\ell}{2}v_\ell^2 + \frac{\lambda_{345}}{2}v_\nu^2 = 0, \quad
m_\nu^2 + \frac{\lambda_\nu}{2}v_\nu^2 + \frac{\lambda_{345}}{2}v_\ell^2 = 0
\end{equation}
with $\lambda_{345}\equiv\lambda_3+\lambda_4+\lambda_5$, fix $\tan\beta\equiv v_\nu/v_\ell$ in terms of the quartic couplings; we treat $\tan\beta$ as a free input in the numerical analysis below.

\section{NUMERICAL ANALYSIS}
\subsection{Neutrino oscillation data}
The texture of Eq.~\eqref{eq:Mnu} is written in terms of four complex Yukawa couplings $a,b,c,d$ and two heavy masses $M_{1,2}$, but most of the phases are unphysical. Using the rephasing to make $a,b$ and $c$ real and positive, and defining $x\equiv c/a>0$, $y\equiv d/b=|y|e^{i\phi_y}$, the light-neutrino mass matrix of Eq.~\eqref{eq:Mnu} takes the form
\begin{equation}
\label{eq:Mnu_reduced}
M_\nu=-A
\begin{pmatrix}
1&0&x\\
0&q&qy\\
x&qy&x^2+qy^2
\end{pmatrix}
\end{equation}
Here, $A=\frac{v_\nu^2a^2}{2M_1},\quad q=\frac{b^2M_1}{a^2M_2}$ and $\phi_y\equiv\arg(d/b)-\arg(c/a)$.

The model is thus controlled by five real parameters, $(A,q,|x|,|y|,\phi_y)$. For a given choice of $(M_1, M_2,\tan\beta)$, the Yukawa couplings are recovered as $a=\sqrt{2M_1A}/v_\nu$, $b=\sqrt{2M_2Aq}/v_\nu$, $c=xa$, $d=yb$, which is how the results of this section are connected to leptogenesis in Section IV(C).

We can obtain the three neutrino mixing angles from 
\begin{equation}
     \begin{aligned}
    \label{eq:17}
    \sin^2\theta_{13} &=|U_{e3}|^2\\ \sin^2\theta_{12} &=\frac{|U_{e2}|^2}{(1-|U_{e3}|^2)}\\ 
    \sin^2\theta_{23} &=\frac{|U_{\mu3}|^2}{(1-|U_{e3}|^2)}
    \end{aligned}
\end{equation}
We can also determine the Dirac phase from the Jarlskog invariant 
\begin{equation}
    \begin{aligned}
        J_{CP} &=\mathrm{Im}[U_{e1}U_{\mu2}U_{e2}^*U_{\mu1}^*]\\
       &=s_{12}c_{12}s_{23}c_{23}s_{13}c_{13}^2\sin\delta_{CP}
    \end{aligned}
\end{equation}
 together with the corresponding $\cos\delta_{CP}$ relation, and the Majorana phase from $\alpha_{21}=2\arg(U_{e1}^*U_{e2})$.
Here, $s_{ij}=\sin{\theta_{ij}}$ and $c_{ij}=\cos{\theta_{ij}}$.\\

For the normal ordering (NO) $m_1=0$, and the condition $(M_\nu)_{e\mu}=0$ becomes $m_2U_{e2}^*U_{\mu2}^*+m_3U_{e3}^*U_{\mu3}^*=0$, which requires
\begin{equation}
\label{eq:19}
\frac{m_2}{m_3}=\frac{|U_{e3}U_{\mu3}|}{|U_{e2}U_{\mu2}|}.
\end{equation}
Scanning the NuFIT~6.1\cite{nufit61} $3\sigma$ intervals of the three angles and all values of $\delta_{CP}$, the right-hand side of equation (~\ref{eq:19}) lies between $0.25$ and $0.46$, whereas the oscillation data give $m_2/m_3=\sqrt{\Delta m^2_{21}/\Delta m^2_{31}}\simeq0.17$--$0.18$. The normal ordering is therefore excluded by the texture itself, independently of the heavy-neutrino masses, so all results below refer to the inverted ordering (IO) with $ m_3=0$ eV.

To perform our numerical analysis, we utilize 3$\sigma$ global-fit data (IO) \cite{esteban2024nufit,nufit61} for neutrino oscillations, as summarized in Table II. These parameter intervals were selected to maintain full agreement with current neutrino oscillation measurements.

\begin{equation}
\begin{aligned}
  & A\in[0.01,0.3]~eV,|x|\in[0.01,3], |y|\in[0.01,3], \\
  & q\in[0.01,3],\phi_y\in[-\pi,\pi]
\end{aligned}
\end{equation}

Across these specified ranges, the input parameters are randomly scanned. Permitted regions are identified by enforcing the 3$\sigma$ constraints on mixing angles and mass-squared differences provided in Table II (for IO).  

Figure~\ref{fig:1} illustrates the correlation between parameters $|x|$ and $|y|$.
\begin{widetext}
\begin{center}
    \includegraphics[width=0.55\textwidth]{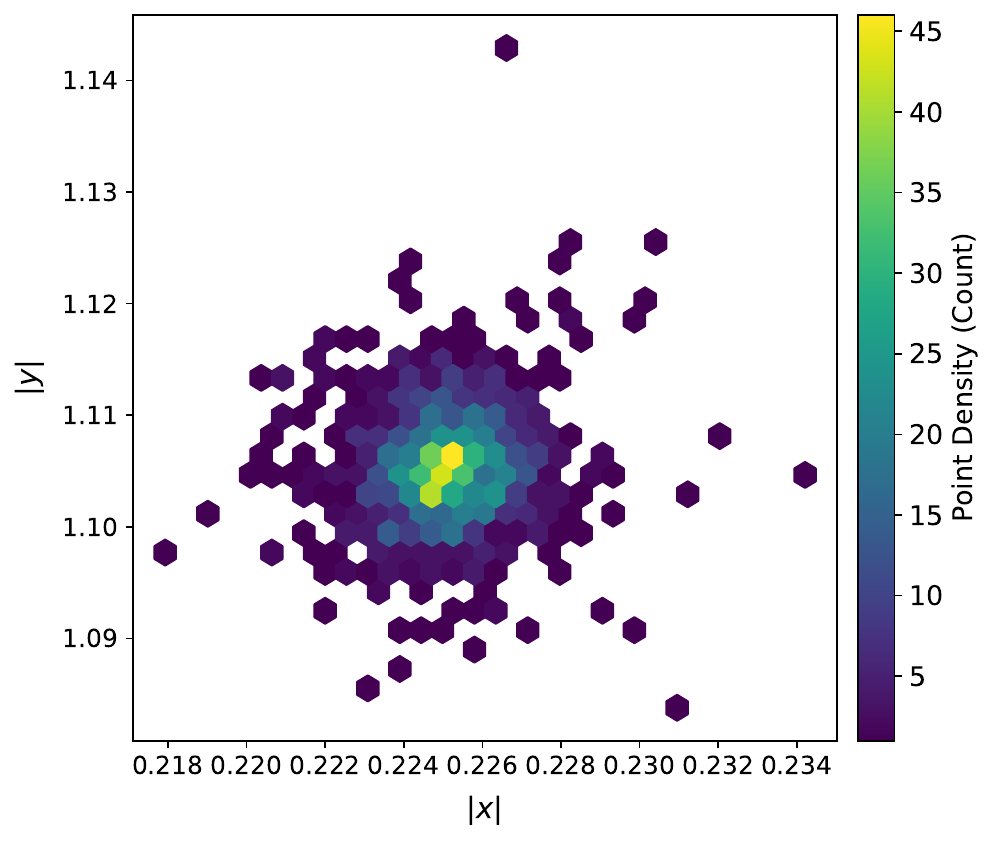}

    \smallskip
    \refstepcounter{figure}
    \textbf{FIG. \thefigure:}
    Correlation between model parameters $|x|$ and $|y|$.
    \label{fig:1}
\end{center}
\end{widetext}

Figure \ref{fig:2} shows the sum of the neutrino masses against the three mixing angles. The prediction satisfies the Planck~2018 bound $\sum m_\nu<0.12eV$ \cite{aghanim2018planck}. Cosmological analyses combining DESI DR2, BAO with CMB data~\cite{adame2025desi,elbers2025constraints} are considerably stronger; they give $\sum m_\nu<0.0642$~eV in $\Lambda$CDM, which is in tension with any inverted-ordering spectrum, but the bound relaxes to $0.163\text{ eV}$ in $w_0w_a$CDM~\cite{elbers2025constraints}, accommodating our findings. Because these constraints depend heavily on the underlying cosmological scenario, we view $\sum m_\nu \simeq 0.098\text{--}0.100\text{ eV}$ as a sharp, falsifiable prediction. Any robust future limit falling below this window would rule out both our model and the inverted ordering entirely.

Figure \ref{fig:3} shows the correlation of the Jarlskog invariant, $J_{CP}$, against the mixing angles(top panels) and the model parameters(bottom panels), with the colour scale giving $\delta_{CP}$.
The $J_{CP}$ is found to be in the range [-0.0349,-0.0322]. In Figure \ref{fig:4}, we have shown the correlation between $J_{CP}$ and $\delta _ {CP}$(colour scale giving point density). 

\begin{widetext}

\begin{figure*}[htbp]
    \centering
        \includegraphics[width=0.77\linewidth]{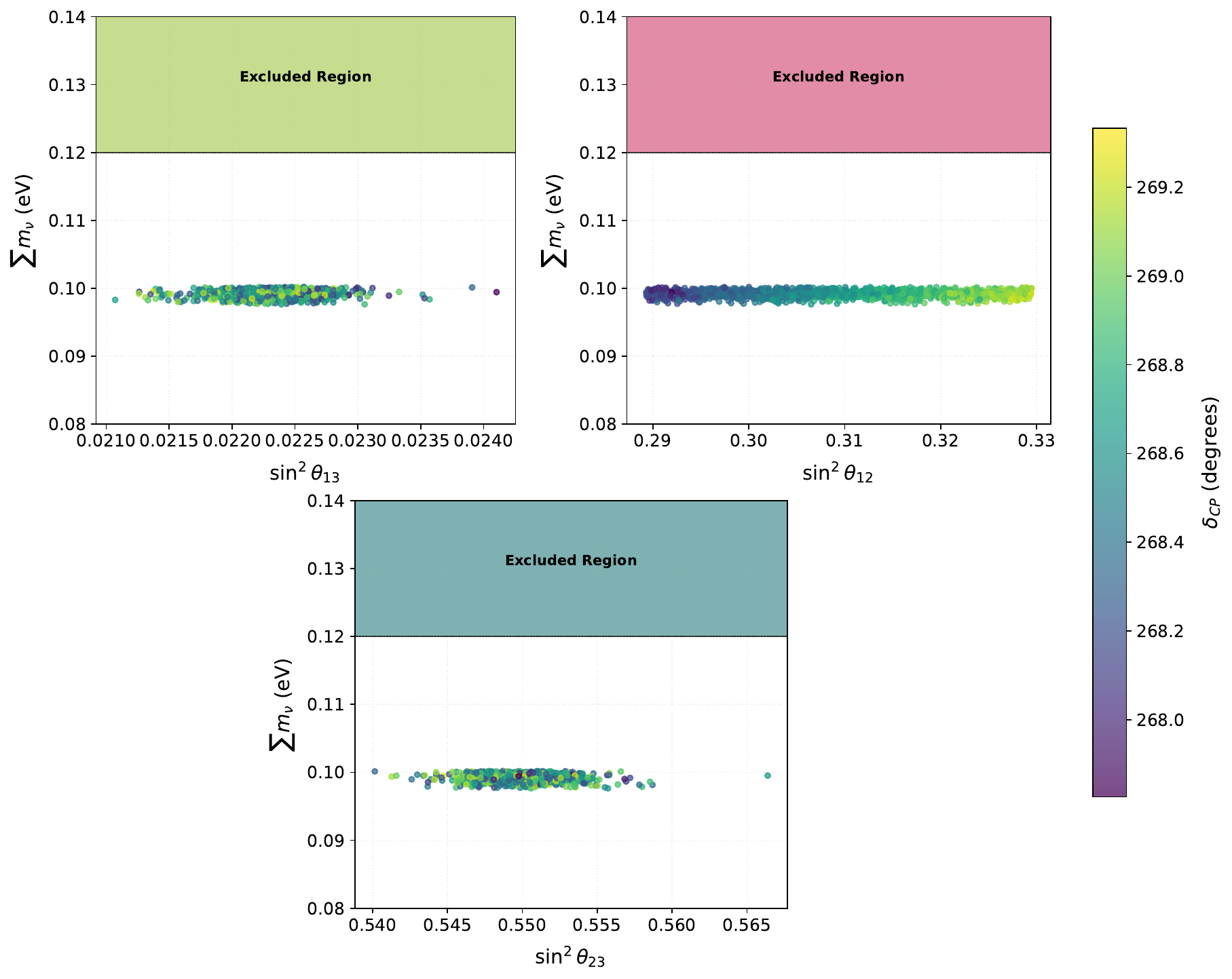}
        \label{fig:sin13}
   
    \caption{Correlations between the mixing angles ($\sin^2\theta_{13}$, $\sin^2\theta_{12}$, $\sin^2\theta_{23}$) and the sum of the neutrino mass $\sum m_{\nu}$.}
    \label{fig:2}
\end{figure*}

\begin{figure*}[htbp]
    \centering
 \includegraphics[width=1\linewidth]{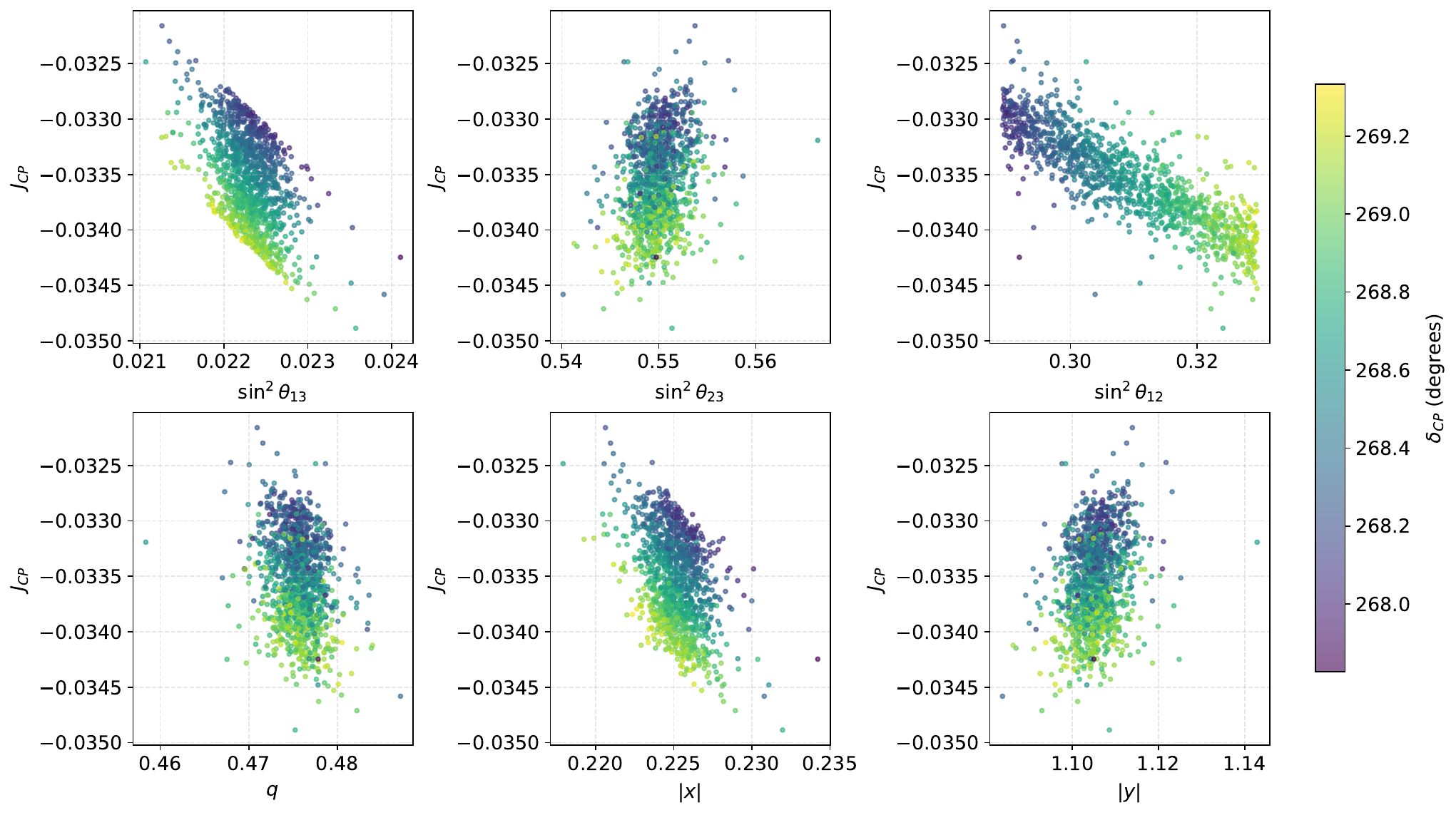}
    
    \caption{Top: Corellation between Jarlskog invariant $J_{CP}$ and $(\sin^2{\theta_{13}}, \sin^2{\theta_{23}}, \sin^2{\theta_{12}})$. Bottom: Correlation between $J_{CP}$ and model parameters $q,|x|,|y|$.}
    \label{fig:3}
\end{figure*}

\begin{figure}[htbp]
 \centering
\includegraphics[width=0.48\linewidth]{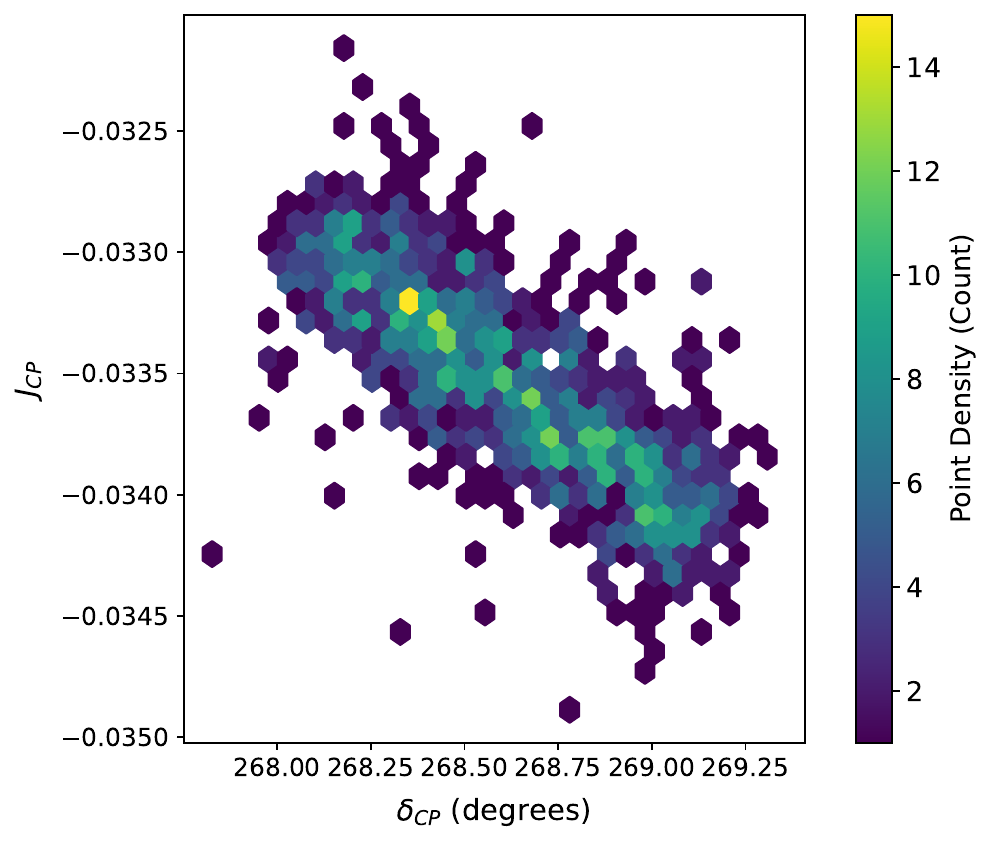}
\caption{Correlation between $J_{cp}$ and $\delta_{cp}$.}
    \label{fig:4}
\end{figure}

\end{widetext}

\begin{table}[htbp]
\label{tab:nufit}
\caption{The NuFIT 6.1 (2025) results\cite{esteban2024nufit,nufit61}.}
\centering
\begin{tabular*}{\linewidth}{l @{\extracolsep{\fill}} cc}
\hline\hline
& \multicolumn{2}{c}{Inverted ordering} \\
\hline
& $\text{bfp} \pm 1\sigma$ & $3\sigma \text{ range}$ \\[6pt]
\hline
$\sin^2 \theta_{12}$ & $0.3088^{+0.0067}_{-0.0066}$ & $0.2893-0.3295$ \\[6pt]
$\sin^2 \theta_{13}$ & $0.02262^{+0.00057}_{-0.00056}$ & $0.02093-0.02441$ \\[6pt]
$\sin^2 \theta_{23}$ & $0.550^{+0.013}_{-0.016}$ & $0.439-0.584$ \\[6pt]
$\dfrac{\Delta m_{21}^2}{10^{-5}\,\text{eV}^2}$ & $7.537^{+0.094}_{-0.10}$ & $7.236-7.822$ \\[6pt]
$\dfrac{\Delta m_{32}^2}{10^{-3}\,\text{eV}^2}$ & $-2.483^{+0.020}_{-0.020}$ & $-2.547--2.421$ \\[6pt]
$\delta_{CP}/^\circ$ & $274^{+22}_{-25}$ & $203-335$ \\[6pt]
\hline\hline
\end{tabular*}
\end{table}

\subsection{Neutrinoless double beta decay ($0\nu\beta\beta$)}
Neutrinoless double beta decay ($0\nu\beta\beta$) is a hypothetical lepton-number-violating nuclear transition described by $(A, Z) \rightarrow (A, Z+2) + 2e^-$\cite{jones2021physics,dolinski2019neutrinoless,bilenky2012neutrinoless}. The observation of this process would establish the Majorana nature of neutrinos and confirm the violation of total lepton number by two units ($\Delta L = 2$).

The decay rate of $0\nu\beta\beta$ is directly proportional to the effective mass $\vert{}m_{\beta \beta}\vert{}$, defined by:
\begin{equation}
m_{\beta \beta} = \left| \sum_{i=1}^{3} U_{ei}^2 m_i \right|.
\end{equation}
Expanding this expression using the PMNS mixing elements $U_{ei}$, the neutrino mass eigenvalues $m_i$, and the Majorana phases leads to:

\begin{equation}
\begin{split}
    \left|m_{\beta \beta}\right| = \Big| & m_1 \cos^2{\theta_{12}}\cos^2{\theta_{13}} \\
    & + m_2 \sin^2{\theta_{12}}\cos^2{\theta_{13}} e^{i\alpha_1} \\
    & + m_3 \sin^2{\theta_{13}} e^{i(\alpha_2 - 2\delta_{CP})} \Big|
\end{split}
\end{equation}

\begin{figure}[htbp]
    \centering
    \includegraphics[width=1\linewidth]{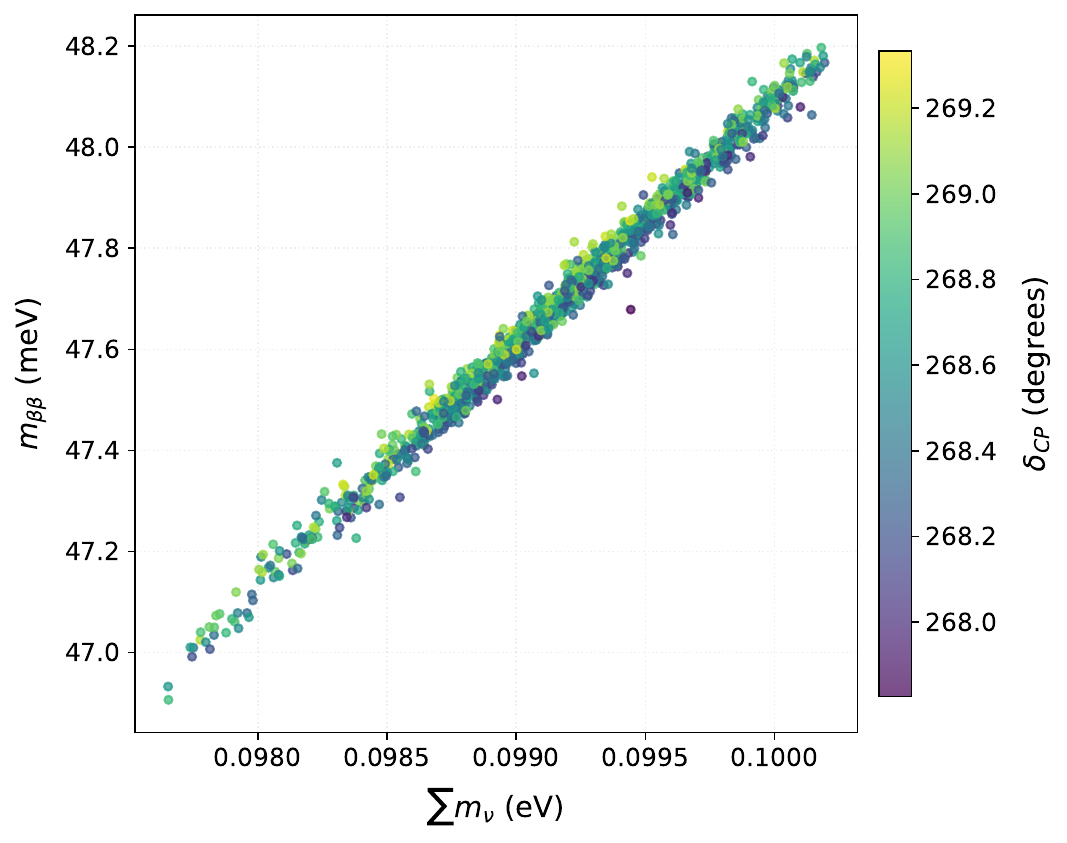}
    \caption{The effective Majorana neutrino mass \(m_{\beta\beta}\) plotted against the total neutrino mass \(\sum m_{\nu}\).}
    \label{fig:5}
\end{figure}

 Figure~\ref{fig:5} shows the correlation between  $m_{\beta\beta}$ and sum of neutrino mass $\sum m_\nu$. The two quantities are almost perfectly correlated. We saw that model predictions lie within KamLAND-Zen~\cite{abe2025search} limit, $m_{\beta\beta}<(28$--$122)$~meV. The projected nEXO~\cite{adhikari2022nexo} sensitivity of $(5$--$20)$~meV at $90\%$~C.L. lies entirely below the predicted range and would therefore test the model decisively.
 
\subsection{Baryon Asymmetry from thermal Leptogenesis}
In thermal leptogenesis, the heavy Majorana neutrinos ($N_i$) are thermally produced in the early Universe following reheating. Their out-of-equilibrium decays into leptonic and Higgs final states ($N_i\to L H_\nu^\dagger$) and ($N_i\to \bar L H_\nu$) generate a net lepton asymmetry through the combined effects of lepton-number violation and CP violation. The CP asymmetry arises from interference between the tree-level decay amplitude and one-loop vertex and self-energy corrections, provided the relevant Yukawa couplings contain physical CP-violating phases~\cite{frossard2013systematic}. As the Universe expands and cools, the decay and inverse-decay processes can depart from thermal equilibrium, allowing the generated asymmetry to survive against washout. The resulting (B-L) asymmetry is subsequently partially converted into a baryon asymmetry by electroweak sphaleron transitions~\cite{d2012sphaleron}, which violate (B+L) while conserving (B-L). Thus, within the seesaw framework, thermal leptogenesis provides a dynamical mechanism for generating the observed baryon asymmetry of the Universe, satisfying the three Sakharov conditions~\cite{sakharov1991violation,sakharov1998violation}: baryon-number violation through sphaleron processes, C and CP violation in the heavy-neutrino interactions, and departure from thermal equilibrium.

The evolution of the heavy-neutrino abundance and the generated \(B-L\) asymmetry can be studied using Boltzmann equations. Defining the evolution parameter \(z=M_1/T\) (here T is the temperature of the thermal bath) and the entropy-normalized abundances \(Y_X=n_X/s\), the unflavoured equations take the form
\begin{equation}
\begin{aligned}
\frac{dY_{N_1}}{dz}
&=-D_1(z)\left(Y_{N_1}-Y_{N_1}^{\mathrm{eq}}\right)\\
\frac{dY_{B-L}}{dz}
&=-\epsilon_1 D_1(z)
\left(Y_{N_1}-Y_{N_1}^{\mathrm{eq}}\right)
-W_1(z)Y_{B-L}
\end{aligned}
\end{equation}
where \(D_1(z)\) describes decays, and \(W_1(z)\) parametrizes the washout processes. These equations provide the basis for determining the final \(B-L\) asymmetry, with the washout term accounting for the relevant inverse-decay and lepton-number-violating scattering processes included in the analysis.

For a hierarchical heavy-neutrino spectrum, $M_1\ll M_{j>1}$, the CP asymmetry in the decays of the lightest heavy Majorana neutrino $N_1$ can be written as
\begin{equation}
\epsilon_1=
\frac{1}{8\pi (Y_\nu^\dagger Y_\nu)_{11}}
\sum_{j>1}
\operatorname{Im}\!\left[
(Y_\nu^\dagger Y_\nu)_{1j}^{\,2}
\right]
f\!\left(x\right),
\end{equation}
where, $x=\frac{M_j^2}{M_1^2}$ and 
\begin{equation}
f(x)=\sqrt{x}\left[
\frac{1}{1-x}+1-(1+x)
\ln\!\left(\frac{1+x}{x}\right)
\right]
\end{equation}

Finally, after the asymmetry freezes out, electroweak sphaleron transitions convert the generated \(B-L\) asymmetry into a baryon asymmetry. Under the appropriate two-Higgs-doublet chemical-equilibrium conditions, the final baryon-to-photon ratio is
\begin{equation}
    Y_B=\frac{8}{23}Y_{B-L}^{\mathrm f},
\qquad
\eta_B\equiv\frac{n_B}{n_\gamma}
\simeq7.04Y_B,
\end{equation}

where \(Y_{B-L}^{\mathrm f}\) denotes the final entropy-normalized \(B-L\) asymmetry and \(\eta_B\) is the baryon-to-photon ratio. The resulting prediction can be compared with the observed baryon asymmetry to constrain the heavy-neutrino masses and neutrino Yukawa parameters of the model.

The observed baryon asymmetry of the universe according to the cosmic microwave background radiation (CMB) data~\cite{de2000flat} and big-bang nucleosynthesis (BBN) data~\cite{cyburt2016big} is
\begin{equation}
\label{eq:27}
    \begin{aligned}
        \eta_B &=(6.08 - 6.16) \times 10^{-10} (CMB)\\
        \eta_B &= (5.99 - 6.14) \times 10^{-10} (BBN)
    \end{aligned}
\end{equation}

To evaluate the baryon asymmetry of the Universe for our model, we employ the Python package \texttt{ULYSSES}~\cite{granelli2021ulysses}. As established in the preceding section, the right-handed neutrino mass matrix and Yukawa couplings naturally satisfy all neutrino oscillation bounds. Using this viable parameter space, we select three representative benchmark points detailed in Table~III for our numerical computation.

The following Yukawa coupling matrices corresponding to these three benchmark points are taken as input for the BAU calculation in the ULYSSES package. 

\begin{equation}
\begin{aligned}
Y_{\nu}(BP1) &= \begin{pmatrix} 
    1.03129 & 0 \\ 
    0 & 0.23405 \\ 
    0.22462 & 0.01100 - 0.23915i 
\end{pmatrix} \\[3ex]
Y_{\nu}(BP2) &= \begin{pmatrix} 
    0.89722 & 0 \\ 
    0 & 0.32164 \\ 
   0.19536 & 0.01521 - 0.32937i 
\end{pmatrix} \\[3ex]
Y_{\nu}(BP3) &= \begin{pmatrix} 
   1.39080 & 0 \\ 
   0 & 0.22314 \\ 
  0.30275 & 0.01071 - 0.22785i 
\end{pmatrix}
\end{aligned}
\end{equation}

\begin{table}[H]
    \centering
    \caption{Benchmark points for unflavoured leptogenesis.}
    \label{tab:leptogenesis_benchmarks}
    \begin{tabular}{ccccc}
        \hline\hline
        \rule{0pt}{3ex} Benchmark & $\tan\beta$ & $M_1$ [GeV] & $M_2$ [GeV] & $M_2/M_1$  \\ [0.5ex]
        \hline
        \rule{0pt}{2.5ex}BP1 & 3.0 & $6.09 \times 10^{13}$ & $6.09 \times 10^{14}$ & 10.0 \\
        BP2 & 1.0 & $6.40 \times 10^{13}$ & $2.56 \times 10^{14}$ & 4.0 \\
        BP3 & 5.0 & $5.92 \times 10^{13}$ & $1.18 \times 10^{15}$ & 20.0 \\ [0.5ex]
        \hline\hline
    \end{tabular}
\end{table}

\begin{figure*}[htbp]
    \centering
    
    \begin{minipage}{0.48\linewidth}
        \centering
        \includegraphics[width=\linewidth]{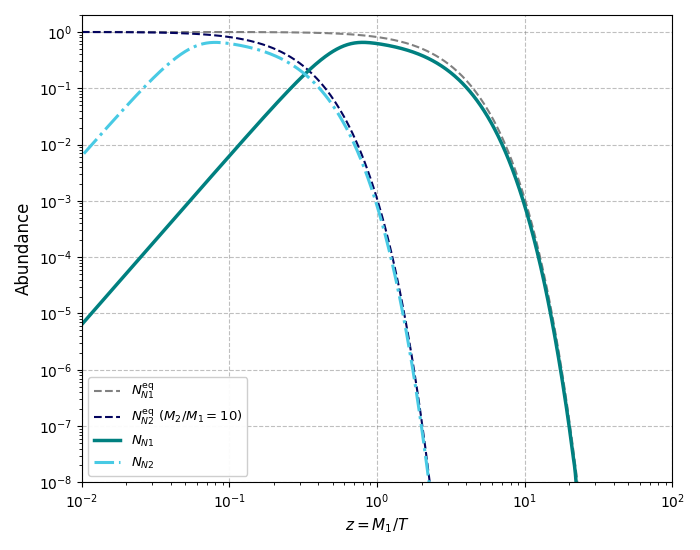}
    \end{minipage}
    \hfill 
    \begin{minipage}{0.48\linewidth}
        \centering
        \includegraphics[width=\linewidth]{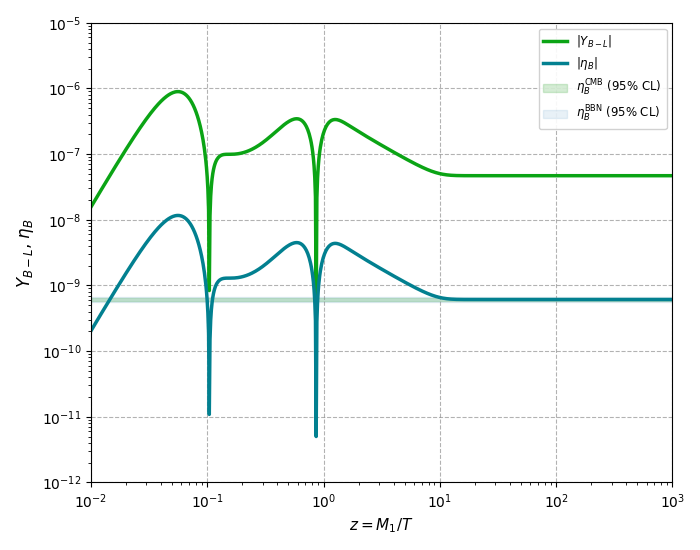}
    \end{minipage}

    \vspace{0.4cm} 
    \begin{minipage}{0.48\linewidth}
        \centering
        \includegraphics[width=\linewidth]{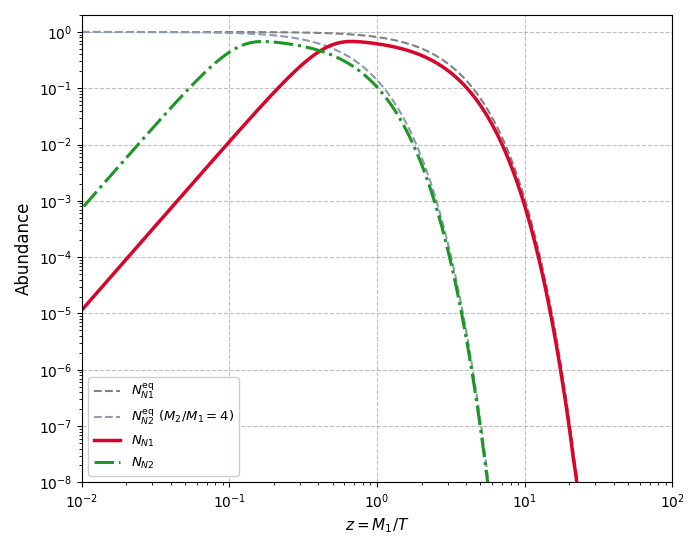}
    \end{minipage}%
    \hfill 
    \begin{minipage}{0.48\linewidth}
        \centering
        \includegraphics[width=\linewidth]{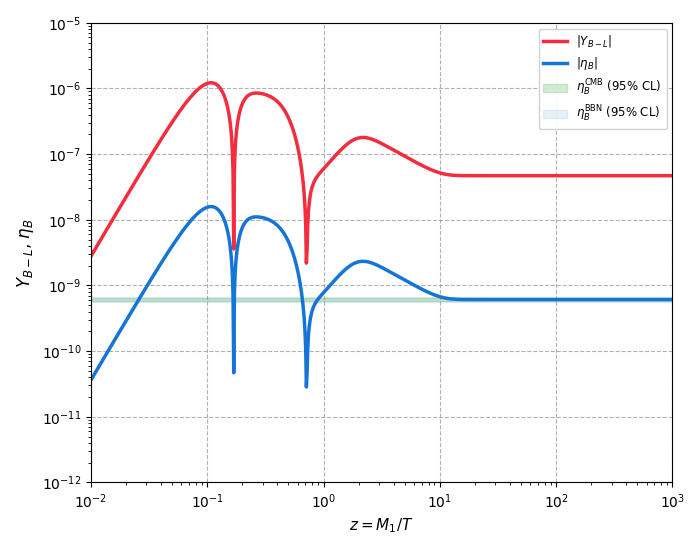}
    \end{minipage}

    \vspace{0.4cm} 

    \begin{minipage}{0.48\linewidth}
        \centering
        \includegraphics[width=\linewidth]{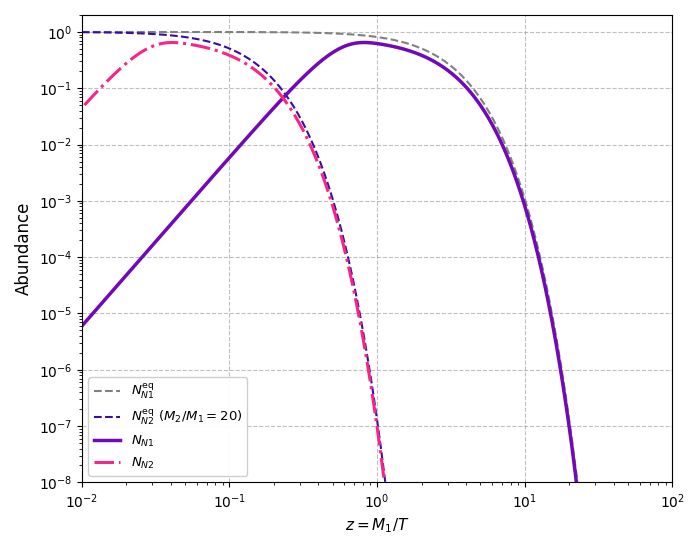} 
    \end{minipage}%
    \hfill 
    \begin{minipage}{0.48\linewidth}
        \centering
        \includegraphics[width=\linewidth]{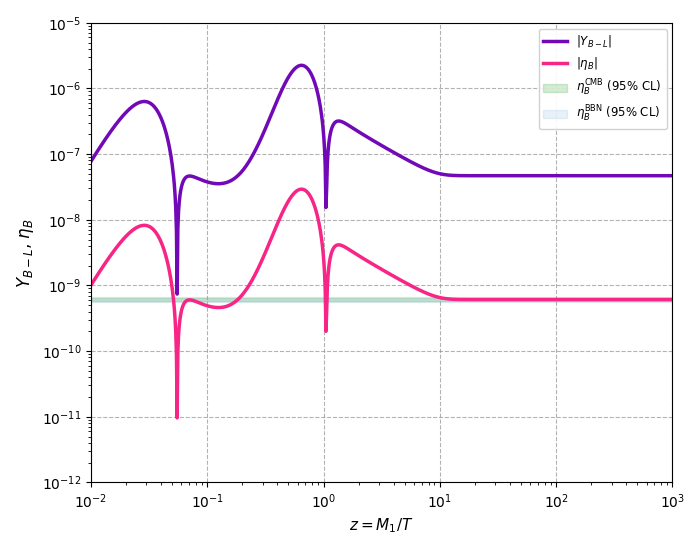} 
    \end{minipage}

    \caption{Evolution of comoving abundances and asymmetries across three benchmark points. The left column shows the comoving abundance of \(N_{1}\) and \(N_{2}\) as a function of z, while the right column tracks the corresponding evolution of lepton and baryon asymmetries. From top to bottom, the rows correspond to benchmark points BP1 (top), BP2 (middle), and BP3 (bottom).}
    \label{fig:BP_All_Six}

\end{figure*}
Figure~\ref{fig:BP_All_Six} illustrates the cosmological evolution of the system: the left panels depict the right-handed neutrino abundances $N_1$ and $N_2$ as functions of the evolution parameter $z = M_1/T$, while the right panels trace the corresponding yields of the lepton asymmetry $Y_{B-L}$ and the baryon asymmetry parameter $\eta_B$ with respect to $z$. As demonstrated in Figure~\ref{fig:BP_All_Six}, the prediction for the baryon asymmetry of the Universe obtained within our model aligns well with the observed value reported in Eq.~(\ref{eq:27}).

\section{Conclusion}

In this work, we have developed a model based on the non-invertible $Z_5^{\mathrm{IN}}$ fusion symmetry to investigate neutrino phenomenology and leptogenesis within a type-I seesaw framework. The non-invertible fusion structure provides an alternative to conventional group-based flavor symmetries and allows us to construct the desired neutrino mass texture without introducing an extended flavon sector or relying on a separate vacuum-alignment mechanism. By embedding the seesaw framework in our two-Higgs-doublet model, we have investigated the implications of the resulting structure for neutrino masses and mixing. Our numerical analysis shows that the model can reproduce the observed neutrino mass-squared differences, with the mixing angles lying within their respective experimentally allowed $3\sigma$ ranges. For inverted mass ordering (IO), the model yields a sum of neutrino masses $\sum m_\nu \simeq [0.098\text{--}0.100]\text{ eV}$, well within the Planck upper limit of $0.12\text{ eV}$. We have calculated the Jarlskog invariant $J_{cp}$ and found to be in the range [-0.0349,-
0.0322]. Also, our model predicts near-maximal CP phase in the range $\delta_{cp}\approx[268^\circ,271^\circ]$. In addition, we have also studied neutrinoless double beta decay, and our predictions for the effective Majorana mass $m_{\beta\beta}$ lie below the current upper bounds from KamLAND-Zen. Furthermore, using the same framework, we computed the baryon asymmetry of the universe via unflavored leptogenesis and found that our model prediction is in close agreement with the observed value of the baryon asymmetry.

\appendix

\nocite{*}

\bibliography{apssamp}

@article{froggatt1979hierarchy,
  title={Hierarchy of quark masses, Cabibbo angles and CP violation},
  author={Froggatt, Colin D and Nielsen, Holger Bech},
  journal={Nuclear Physics B},
  volume={147},
  number={3-4},
  pages={277--298},
  year={1979},
  publisher={Elsevier}
}

@article{adame2025desi,
  title={DESI 2024 VI: cosmological constraints from the measurements of baryon acoustic oscillations},
  author={Adame, Armand G and Aguilar, J and Ahlen, Steven and Alam, Shadab and Alexander, David M and Alvarez, M and Alves, Ot{\'a}vio and Anand, Abhijeet and Andrade, Uendert and Armengaud, Eric and others},
  journal={Journal of Cosmology and Astroparticle Physics},
  volume={2025},
  number={02},
  pages={021},
  year={2025},
  publisher={IOP Publishing}
}

@article{elbers2025constraints,
  title={Constraints on neutrino physics from DESI DR2 BAO and DR1 full shape},
  author={Elbers, Willem and Avil{\'e}s, Alejandro and Noriega, HE and Chebat, D and Menegas, Alexander and Frenk, Carlos S and Garcia-Quintero, C and Gonzalez, D and Ishak, Mustapha and Lahav, O and others},
  journal={Physical Review D},
  volume={112},
  number={8},
  pages={083513},
  year={2025},
  publisher={APS}
}

@article{ma2006neutrino,
  title={Neutrino mass matrix from S4 symmetry},
  author={Ma, Ernest},
  journal={Physics Letters B},
  volume={632},
  number={2-3},
  pages={352--356},
  year={2006},
  publisher={Elsevier}
}

@article{altarelli2009revisiting,
  title={Revisiting bimaximal neutrino mixing in a model with S 4 discrete symmetry},
  author={Altarelli, Guido and Feruglio, Ferruccio and Merlo, Luca},
  journal={Journal of High Energy Physics},
  volume={2009},
  number={05},
  pages={020--020},
  year={2009}
}

@article{altarelli2010discrete,
  title={Discrete flavor symmetries and models of neutrino mixing},
  author={Altarelli, Guido and Feruglio, Ferruccio},
  journal={Reviews of Modern Physics},
  volume={82},
  number={3},
  pages={2701--2729},
  year={2010},
  publisher={APS}
}

@article{altarelli2006tri,
  title={Tri-bimaximal neutrino mixing, A4 and the modular symmetry},
  author={Altarelli, Guido and Feruglio, Ferruccio},
  journal={Nuclear Physics B},
  volume={741},
  number={1-2},
  pages={215--235},
  year={2006},
  publisher={Elsevier}
}

@article{babu2003underlying,
  title={Underlying A4 symmetry for the neutrino mass matrix and the quark mixing matrix},
  author={Babu, Kaladi S and Ma, Ernest and Valle, JWF},
  journal={Physics Letters B},
  volume={552},
  number={3-4},
  pages={207--213},
  year={2003},
  publisher={Elsevier}
}

@article{borah2026gravitational,
  author={Borah, Pulakesh and Das, Mrinal Kumar},
  title={Gravitational Wave in an {$A_4 \otimes Z_4$} Neutrino Model with Generalised {CP}},
  journal={arXiv preprint arXiv:2609.07661},
  year={2026}
}

@article{das2020phenomenological,
  title={Phenomenological study of neutrino mass, dark matter and baryogenesis within the framework of minimal extended seesaw},
  author={Das, Pritam and Das, Mrinal Kumar and Khan, Najimuddin},
  journal={Journal of High Energy Physics},
  volume={2020},
  number={3},
  pages={18},
  year={2020},
  publisher={Springer}
}

@article{pathak2026matter,
  title={Matter-antimatter asymmetry in minimal inverse seesaw framework with A 4 modular symmetry},
  author={Pathak, Gourab and Das, Mrinal Kumar},
  journal={Journal of Physics G: Nuclear and Particle Physics},
  volume={53},
  number={2},
  pages={025004},
  year={2026},
  publisher={IOP Publishing}
}

@article{kumar2025neutrino,
  title={Neutrino Textures from Modular $ A\_4 $ Left--Right Symmetry: Experimental Signatures at DUNE and T2HK in the Post-JUNO Era},
  author={Kumar, Bhabana and Dutta, Debajyoti and Das, Mrinal Kumar},
  journal={arXiv preprint arXiv:2512.22020},
  year={2025}
}

@article{kobayashi2025radiative,
  title={Radiative neutrino mass models from non-invertible selection rules},
  author={Kobayashi, Tatsuo and Okada, Hiroshi and Otsuka, Hajime},
  journal={Journal of High Energy Physics},
  volume={2025},
  number={12},
  pages={111},
  year={2025},
  publisher={Springer}
}

@article{everett2009icosahedral,
  title={Icosahedral (A 5) family symmetry and the golden ratio prediction for solar neutrino mixing},
  author={Everett, Lisa L and Stuart, Alexander J},
  journal={Physical Review D—Particles, Fields, Gravitation, and Cosmology},
  volume={79},
  number={8},
  pages={085005},
  year={2009},
  publisher={APS}
}

@article{ding2012golden,
  title={Golden ratio neutrino mixing and A5 flavor symmetry},
  author={Ding, Gui-Jun and Everett, Lisa L and Stuart, Alexander J},
  journal={Nuclear Physics B},
  volume={857},
  number={3},
  pages={219--253},
  year={2012},
  publisher={Elsevier}
}

@incollection{feruglio2019neutrino,
  title={Are neutrino masses modular forms?},
  author={Feruglio, Ferruccio},
  booktitle={From My Vast Repertoire… Guido Altarelli’s Legacy},
  pages={227--266},
  year={2019},
  publisher={World Scientific}
}

@article{jiang2025texture,
  title={Texture-zeros in minimal seesaw from noninvertible symmetry fusion rules},
  author={Jiang, Zheng and Qu, Bu-Yao and Ding, Gui-Jun},
  journal={Physical Review D},
  volume={112},
  number={11},
  pages={115029},
  year={2025},
  publisher={APS}
}

@article{brdar2019type,
  title={Type I seesaw mechanism as the common origin of neutrino mass, baryon asymmetry, and the electroweak scale},
  author={Brdar, Vedran and Helmboldt, Alexander J and Iwamoto, Sho and Schmitz, Kai},
  journal={Physical Review D},
  volume={100},
  number={7},
  pages={075029},
  year={2019},
  publisher={APS}
}

@article{minkowski1977mu,
  title={$\mu$→ e$\gamma$ at a rate of one out of 109 muon decays?},
  author={Minkowski, Peter},
  journal={Physics Letters B},
  volume={67},
  number={4},
  pages={421--428},
  year={1977},
  publisher={Elsevier}
}

@article{mohapatra1980neutrino,
  title={Neutrino mass and spontaneous parity nonconservation},
  author={Mohapatra, Rabindra N and Senjanovi{\'c}, Goran},
  journal={Physical Review Letters},
  volume={44},
  number={14},
  pages={912},
  year={1980},
  publisher={APS}
}

@article{buchmuller2005leptogenesis,
  title={Leptogenesis as the origin of matter},
  author={Buchm{\"u}ller, Wilfried and Peccei, Roberto D and Yanagida, Tsutomu},
  journal={Annu. Rev. Nucl. Part. Sci.},
  volume={55},
  number={1},
  pages={311--355},
  year={2005},
  publisher={Annual Reviews}
}

@article{giudice2004towards,
  title={Towards a complete theory of thermal leptogenesis in the SM and MSSM},
  author={Giudice, Gian Francesco and Notari, A and Raidal, M and Riotto, A and Strumia, Alessandro},
  journal={Nuclear Physics B},
  volume={685},
  number={1-3},
  pages={89--149},
  year={2004},
  publisher={Elsevier}
}

@article{vives2006flavor,
  title={Flavor dependence of CP asymmetries and thermal leptogenesis with strong right-handed neutrino mass hierarchy},
  author={Vives, O},
  journal={Physical Review D—Particles, Fields, Gravitation, and Cosmology},
  volume={73},
  number={7},
  pages={073006},
  year={2006},
  publisher={APS}
}

@incollection{sakharov1998violation,
  title={Violation of CP-invariance, C-asymmetry, and baryon asymmetry of the Universe},
  author={Sakharov, Andrei D},
  booktitle={In The Intermissions… Collected Works on Research into the Essentials of Theoretical Physics in Russian Federal Nuclear Center, Arzamas-16},
  pages={84--87},
  year={1998},
  publisher={World Scientific}
}

@article{sakharov1991violation,
  title={Violation of CP invariance, C asymmetry, and baryon asymmetry of the universe},
  author={Sakharov, Andrei D},
  journal={Sov. Phys. Usp},
  volume={34},
  number={5},
  pages={392--393},
  year={1991}
}

@article{Zeldovich:1974uw,
    author = "Zeldovich, Ya. B. and Kobzarev, I. Yu. and Okun, L. B.",
    title = "{Cosmological Consequences of the Spontaneous Breakdown of Discrete Symmetry}",
    reportNumber = "SLAC-TRANS-0165, IPM-MOSCOW-15",
    journal = "Zh. Eksp. Teor. Fiz.",
    volume = "67",
    pages = "3--11",
    year = "1974"
}

@article{chauhan2024phenomenology,
  title={Phenomenology of lepton masses and mixing with discrete flavor symmetries},
  author={Chauhan, Garv and Dev, PS Bhupal and Dubovyk, Ievgen and Dziewit, Bartosz and Flieger, Wojciech and Grzanka, Krzysztof and Gluza, Janusz and Karmakar, Biswajit and Zi{\fontencoding{T1}\selectfont\k{e}}ba, Szymon},
  journal={Progress in Particle and Nuclear Physics},
  volume={138},
  pages={104126},
  year={2024},
  publisher={Elsevier}
}

@article{jones2021physics,
  title={The physics of neutrinoless double beta decay: A primer},
  author={Jones, Benjamin JP},
  journal={arXiv preprint arXiv:2108.09364},
  year={2021}
}

@article{dolinski2019neutrinoless,
  title={Neutrinoless double-beta decay: status and prospects},
  author={Dolinski, Michelle J and Poon, Alan WP and Rodejohann, Werner},
  journal={Annual Review of Nuclear and Particle Science},
  volume={69},
  number={1},
  pages={219--251},
  year={2019},
  publisher={Annual Reviews}
}

@article{bilenky2012neutrinoless,
  title={Neutrinoless double-beta decay: A brief review},
  author={Bilenky, SM and Giunti, Carlo},
  journal={Modern Physics Letters A},
  volume={27},
  number={13},
  pages={1230015},
  year={2012},
  publisher={World Scientific}
}

@article{gomez2011sense,
  title={Sense and sensitivity of double beta decay experiments},
  author={G{\'o}mez-Cadenas, Juan Jos{\'e} and Mart{\'\i}n-Albo, Justo and Sorel, Michel and Ferrario, Paola and Monrabal, Francesc and Munoz, J and Novella, Pau and Poves, Alfredo},
  journal={Journal of Cosmology and Astroparticle Physics},
  volume={2011},
  number={06},
  pages={007--007},
  year={2011}
}

@article{frossard2013systematic,
  title={Systematic approach to thermal leptogenesis},
  author={Frossard, Tibor and Garny, Mathias and Hohenegger, Andreas and Kartavtsev, Alexander and Mitrouskas, David},
  journal={Physical Review D—Particles, Fields, Gravitation, and Cosmology},
  volume={87},
  number={8},
  pages={085009},
  year={2013},
  publisher={APS}
}

@article{d2012sphaleron,
  title={The sphaleron rate through the electroweak cross-over},
  author={D’Onofrio, Michela and Rummukainen, Kari and Tranberg, Anders},
  journal={Journal of High Energy Physics},
  volume={2012},
  number={8},
  pages={123},
  year={2012},
  publisher={Springer}
}

@article{ma2001softly,
  title={Softly broken A 4 symmetry for nearly degenerate neutrino masses},
  author={Ma, Ernest and Rajasekaran, G},
  journal={Physical Review D},
  volume={64},
  number={11},
  pages={113012},
  year={2001},
  publisher={APS}
}

@article{aghanim2018planck,
  title={Planck 2018 results},
  author={Aghanim, N and Akrami, Y and Ashdown, M and Aumont, J and Baccigalupi, C and Ballardini, M and Banday, AJ and Barreiro, RB and Bartolo, N and Basak, S and others},
  journal={VI. Cosmological parameters},
  volume={7},
  pages={1807--06209},
  year={2018}
}

@article{adhikari2022nexo,
  title={nEXO: neutrinoless double beta decay search beyond 1028 year half-life sensitivity},
  author={Adhikari, Govinda and Al Kharusi, S and Angelico, E and Anton, G and Arnquist, IJ and Badhrees, I and Bane, J and Belov, V and Bernard, EP and Bhatta, T and others},
  journal={Journal of Physics G: Nuclear and Particle Physics},
  volume={49},
  number={1},
  pages={015104},
  year={2022},
  publisher={IOP Publishing}
}

@article{abe2025search,
  title={Search for Majorana neutrinos with the complete KamLAND-Zen dataset},
  author={Abe, S and Araki, T and Chiba, K and Eda, T and Eizuka, M and Funahashi, Y and Furuto, A and Gando, A and Gando, Y and Goto, S and others},
  journal={Physical Review Letters},
  volume={135},
  number={26},
  pages={262501},
  year={2025},
  publisher={APS}
}

@article{esteban2024nufit,
  title={NuFit-6.0: updated global analysis of three-flavor neutrino oscillations},
  author={Esteban, Ivan and Gonzalez-Garcia, Ma Concepci{\'o}n and Maltoni, Michele and Martinez-Soler, Ivan and Pinheiro, Jo{\~a}o Paulo and Schwetz, Thomas},
  journal={Journal of High Energy Physics},
  volume={2024},
  number={12},
  pages={216},
  year={2024},
  publisher={Springer}
}

@misc{nufit61,
  author = {{NuFIT Collaboration}},
  title = {{NuFIT 6.1 data release}},
  year = {2025},
  url = {http://www.nu-fit.org/?q=node/309}
}

@article{de2000flat,
  title={A flat Universe from high-resolution maps of the cosmic microwave background radiation},
  author={de Bernardis, Pea and Ade, Peter AR and Bock, James J and Bond, JR and Borrill, J and Boscaleri, A and Coble, K and Crill, BP and De Gasperis, G and Farese, PC and others},
  journal={Nature},
  volume={404},
  number={6781},
  pages={955--959},
  year={2000},
  publisher={Nature Publishing Group UK London}
}

@article{cyburt2016big,
  title={Big bang nucleosynthesis: Present status},
  author={Cyburt, Richard H and Fields, Brian D and Olive, Keith A and Yeh, Tsung-Han},
  journal={Reviews of Modern Physics},
  volume={88},
  number={1},
  pages={015004},
  year={2016},
  publisher={APS}
}

@article{granelli2021ulysses,
  title={ULYSSES: Universal LeptogeneSiS equation solver},
  author={Granelli, Alessandro and Moffat, Kristian and Perez-Gonzalez, Yuber F and Schulz, Holger and Turner, Jessica},
  journal={Computer Physics Communications},
  volume={262},
  pages={107813},
  year={2021},
  publisher={Elsevier}
}

\end{document}